\documentclass[twoside,twocolumn]{elsart}

\usepackage{graphicx}
\graphicspath{{${HOME}/fig/publi/muon/}}

\begin{document}
\begin{frontmatter}



\title{Evidence for an antiferromagnetic component in the magnetic structure of ZrZn$_2$} 

\author[CEA]{P. Dalmas de R\'eotier}
\author[CEA]{G. Lapertot}
\author[CEA]{A. Yaouanc}
\author[IRI]{P.C.M. Gubbens}
\author[IRI]{S. Sakarya}
\author[PSI]{A. Amato}

\address[CEA]{Commissariat \`a l'Energie Atomique, D\'epartement de Recherche
Fondamentale sur la Mati\`ere Condens\'ee, SPSMS,
F-38054 Grenoble cedex 9, France}

\address[IRI]{Department of Radiation, Radionuclides \& Reactors,
Faculty of Applied Sciences,
Delft University of Technology, 2629 JB Delft, The Netherlands}

\address[PSI]{Laboratory for Muon-Spin Spectroscopy, Paul Scherrer Institute, 
5232 Villigen-PSI, Switzerland}

\begin{abstract}

Zero-field muon spin rotation experiments provide evidence for an antiferromagnetic 
component in the magnetic structure of the intermetallics ZrZn$_2$. 

\end{abstract}

\end{frontmatter}


The discovery of superconductivity below 1 K within a limited pressure range 
in UGe$_2$ provided an unanticipated example of the coexistence of 
superconductivity and 
ferromagnetism \cite{Saxena00}. Later on, resistivity measurements
indicated the onset of superconductivity in ZrZn$_2$ at ambient pressure
below $\sim$ 0.3~K, {\sl i.e.}
deep in the ferromagnetic ordered state of this $d$-electron compound 
\cite{Pfleiderer01}. 
The electronic pairing mechanism needed for superconductivity is believed to 
be of magnetic origin for both materials.

It is important to ascertain the magnetic structure of these two compounds 
because a
Cooper's pairing mechanism in compounds with a spin density wave has been 
described in the 
literature \cite{Watanabe02}. The ferromagnetic state of UGe$_2$ 
is in fact not 
simple since bulk measurements provide evidence for a magnetic transition within the 
ferromagnet state. Its origin is unknown, despite a number of neutron diffraction investigations. 
The general belief that the ground state of ZrZn$_2$ is ferromagnetic below $T_{\rm C} \simeq$  28~K 
rests on measurements performed more than 40 years ago 
\cite{Matthias58,Pickart64}. 
A magnetic structure which would be characterised in neutron diffraction by weak
magnetic satellites in the vicinity of a main ferromagnetic Bragg reflection is not excluded experimentally.
The satellites would mean that the magnetic structure of ZrZn$_2$ contains an antiferromagnetic (AFM) 
component. Since the saturation moment is relatively small and the magnetic structure the 
compound may adopt is unknown, the muon spin rotation ($\mu$SR) technique is the most convenient method 
to determine whether an AFM component is at all possible. 
Here we report the results of measurements with this technique, which demonstrate the existence of an AFM
component in the magnetic structure of ZrZn$_2$.

The ZrZn$_2$ powder sample was made by slow cooling of an optimized alloy
of Zr and Zn with atomic ratio
Zr:Zn = 1:2.006 in order 
to avoid the formation of surrounding secondary phases (ZrZn$_3$ and ZrZn). 
The starting materials were zone refined zirconium and zinc of 6N purity.
To prevent the contamination of ZrZn$_2$ by aluminium if an alumina crucible was used, the mixture
of zirconium and zinc was introduced in a home made yttria crucible which was encapsulated in a 
tantalum crucible to hold the zinc vapor pressure. No further annealing was 
performed on this sample.

The analysis of x-ray diffraction patterns reveals no parasitic phases. 
They are consistent 
with the C15 cubic Laves structure.
Figure 1 shows the magnetisation measured at low temperature in a 
field of 0.1 T. The magnetic ordering temperature was determined from 
Arrott's plot (not shown) to be close to 27~K.
\begin{figure}
\centering
\includegraphics[width=0.45\textwidth]{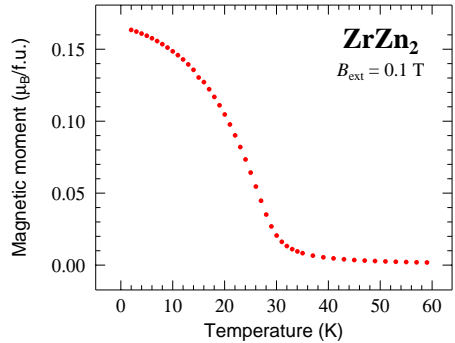}
\caption{Magnetisation versus temperature measured in an external
field of $B_{\rm ext}$ = 0.1 T on the 
sample of ZrZn$_2$ used for the $\mu$SR experiment. 
}
\label{Gerard}
\end{figure}

The zero-field $\mu$SR measurements were performed at the $\pi$M3 beamline of the Swiss muon 
source (Paul Scherrer Institute, Villigen, Switzerland) from 1.7 up to 66 K. 

The $\mu$SR technique consists of implanting polarised 
(along direction $Z$) muons into a specimen and monitoring  $P_Z^{\rm exp}(t)$ 
which provides a measure of the evolution of the muon ensemble polarisation projected onto direction 
$Z$; for an introduction to this technique, see e.g. 
Refs. \cite{Schenck95,Karlsson95,Dalmas97}. The quantity 
actually measured is the so-called asymmetry corresponding to $a_0 P_Z^{\rm exp}(t)$, 
where $a_0$ $\simeq$ 0.24.

For a paramagnet, $P_Z^{\rm exp}(t)$ tracks the dynamics of the magnetic field 
at the muon site, reflecting the dynamics of the electronic moments.
In the fast fluctuation or motional narrowing limit, $P_Z^{\rm exp}(t)$ 
takes the exponential form $\exp(-\lambda_Z t)$ characterised by a
relaxation rate $\lambda_Z$. The measurements above 
$T_{\rm C}$ are fully consistent with this expectation. $\lambda_Z$ is found to be smaller than 
0.01 $\mu {\rm s}^{-1}$. The thermal dependence of $\lambda_Z$ above $T_{\rm C}$
has not been studied in details because its behaviour is not easily related to the magnetic structure 
of the compound under investigation.

\begin{figure}
\centering
\includegraphics[width=0.45\textwidth]{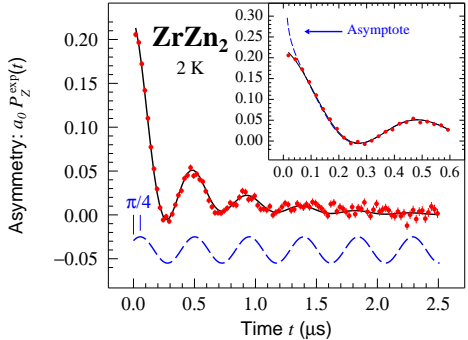}
\caption{A zero-field $\mu$SR spectrum recorded for ZrZn$_2$ at 2 K, i.e.
deep in the ordered magnetic state. The full line is the result of a fit using 
Eq. \ref{fit} which includes a Bessel function
as explained in the text. The inset presents the results of the 
fit using the long-time asymptotic limit of the Bessel function. This limit 
is actually valid in most of the time range.
That a phase shift of $-\pi/4$ can be detected, is shown in the main panel 
where a cosine function shifted by $-\pi/4$ is plotted. Note that the 
extremum displayed by this function 
at small times is not seen in the measured spectrum.}
\label{spectrum}
\end{figure}

For a measurement performed in the long-range magnetically ordered state of a powder sample such as 
ours, $P_Z^{\rm exp}(t)$ is expected to be the sum of a damped oscillation for two third of the 
asymmetry and a simple exponential relaxation function for one third. As shown in Fig.\ \ref{spectrum},   
these expectations are borne out by the experiment. The properties of the oscillating 
component contains the information we are looking for.
The fit of the spectrum is done using the following expression for $P_Z^{\rm exp}(t)$:
\begin{eqnarray}
P_Z^{\rm exp}(t) & = & {p_1}\exp \left ( - \lambda_X t \right )
J_0 \left (\gamma_\mu B_{\rm max} t \right ) \cr
& + &  {p_2} \exp \left ( - \lambda_Z t \right ),
\label{fit}
\end{eqnarray} 
with $p_1 + p_2 =1$.
The damped oscillation is modeled by the product of 
two functions: a zeroth-order Bessel function $J_0$ and an exponential 
function.
$J_0$ is the cosine-Fourier transform of the field distribution at the muon
site resulting from a simple incommensurate magnetic structure 
\cite{Overhauser60}. However, it is possible to generate a similar distribution 
out of a complex magnetic structure; see for example Fig.~10 in Ref.~\cite{Chiba92}.
Anyway, $J_0$-like oscillation indicates clearly that 
ZrZn$_2$ cannot be a simple ferromagnet. 
The exponential envelop of the oscillations is thought to account for the finite coherence length, {\sl i.e.}
the disorder of the magnetic structure. $\gamma_\mu$ is the muon gyromagnetic 
ratio ($\gamma_\mu$ = 851.6 Mrad~s$^{-1}$~T$^{-1}$), $B_{\rm max}$ the maximum 
value of the magnetic field at the muon site and $\lambda_Z$ the 
spin-lattice relaxation rate. Numerically,
the fit of the spectrum in Fig.\ \ref{spectrum} gives $B_{\rm max}= 16.5 \,(1)$ mT, 
$\lambda_X = 1.42 \, (9) \, \mu{\rm s}^{-1}$ and $\lambda_Z= 1.63\, (11)\, \mu{\rm s}^{-1}$. 
We find $p_1/p_2 \sim 2.7$ instead of the expected value of $2$. This discrepancy 
is a signature of a relatively large texture for our sample, which is not
surprising since it is made of relatively large crystallites.
Spectra were recorded for increasing temperatures up to $T_{\rm C}$. 
They are all 
well described by Eq.\ \ref{fit}.  

To reach a reliable conclusion concerning the possible presence of an AFM component
in the magnetic structure of ZrZn$_2$, we need to determine whether the description of the 
wiggles displayed by $P_Z^{\rm exp}(t)$ with the zeroth-order Bessel function is safe.
We have done a fit of the measured spectrum substituting in Eq.\ \ref{fit}
$J_0(\gamma_\mu B_{\rm max} t)$ with the cosine function 
$\cos (\gamma_\mu B_{\rm loc} t +  \varphi)$. We 
derive $B_{\rm loc} = 16.0 \, (2)$ mT. Remarkably, $  \varphi$  = $-34. 5$ (2.2) degrees. 
The phase shift $\varphi$ is therefore relatively large whereas to get a physically 
grounded cosine fit, $\varphi$ should be negligible, reflecting only the inherent
error of the experimental determination of the time origin. We 
have checked that point by recording zero-field 
spectra on the intermetallics PrRu$_2$Si$_2$ which is a collinear ferromagnet 
below 14~K \cite{Mulders97} and which is characterised by $B_{\rm loc}$
of the same order of magnitude as ZrZn$_2$.
As expected, the observed oscillation is 
well accounted for by a cosine function with a negligible phase shift.

The fact that the fit with a shifted cosine function provides a description 
of the spectrum with the exception of the short times, is easily understood 
if we recall that $J_0(x)$ is well approximated by the asymptote
$({2\over \pi x})^{1\over 2} \cos(x-{\pi\over 4})$ valid if $x$ is not 
too small.   

In conclusion, the $J_0(x)$ Bessel function provides a reliable description of the measured 
$\mu$SR oscillation in the ordered state of ZrZn$_2$. This means that this compound is not a simple 
ferromagnet and that an AFM component is therefore present.
However, our measurements do not resolve the magnetic structure, as no 
information on propagation wavevectors could be extracted from the $\mu$SR 
data. 
Since the superconductivity of ZrZn$_2$ is thought to be of magnetic origin,
a proper characterisation of its magnetic structure is important.
This could be achieved using neutron scattering experiments. 
Among the possibilities for the magnetic structure, a well known example 
for a compound with a spiral structure is provided by 
CeAl$_2$ \cite{Forgan90}. Interestingly, CeAl$_2$ shares the cubic C15 
crystal structure with ZrZn$_2$. However, the length of the wavevector is 
substantial for CeAl$_2$
whereas, if it exists, it is probably small for ZrZn$_2$. 

We are grateful to  T. Jarlborg and S. Pouget for useful discussions.
Part of this work was performed at the
S$\mu$S, Paul Scherrer Institute, Villigen, Switzerland. 

\bibliographystyle{elsart-num}
\bibliography{ZrZn2_structure}
\end{document}